%% file: main.tex
\documentclass{article}
\usepackage{spconf,amsmath,graphicx}
\usepackage{amsmath,amsfonts,amssymb,pifont}
\usepackage{adjustbox}
\usepackage{comment}
\usepackage[para]{footmisc}
\usepackage{xparse}

\usepackage{multirow,makecell}

\title{Audio Retrieval with WavText5K and CLAP Training \vspace{-7pt}}

\name{Soham Deshmukh, Benjamin Elizalde, Huaming Wang \vspace{-8.5pt}}
\address{Microsoft \\
\{sdeshmukh, benjaminm, huawang\}@microsoft.com \vspace{-8pt}}

\begin{document}
%
\maketitle
\input{texfiles/abstract2}
\input{texfiles/introduction2}
\input{texfiles/newdataset}
\input{texfiles/architecture}
\input{texfiles/experiments}
\input{texfiles/results}
\input{texfiles/conclusions}

\pagebreak

\bibliographystyle{IEEEbib}
\bibliography{refs}

\clearpage
\appendix

\end{document}

%% file: texfiles/abstract2.tex
\begin{abstract}
\vspace{-0.01in}

Audio-Text retrieval takes a natural language query to retrieve relevant audio files in a database. Conversely, Text-Audio retrieval takes an audio file as a query to retrieve relevant natural language descriptions. Most of the literature train retrieval systems with one audio captioning dataset, but evaluating the benefit of training with multiple datasets is underexplored. Moreover, retrieval systems have to learn the alignment between elaborated sentences describing audio content of variable length ranging from a few seconds to several minutes. In this work, we propose a new collection of web audio-text pairs and a new framework for retrieval. First, we provide a new collection of about five thousand web audio-text pairs that we refer to as WavText5K. When used to train our retrieval system, WavText5K improved performance more than other audio captioning datasets. Second, our framework learns to connect language and audio content by using a text encoder, two audio encoders, and a contrastive learning objective. Combining both audio encoders helps to process variable length audio. The two contributions beat state of the art performance for AudioCaps and Clotho on Text-Audio retrieval by a relative 2\% and 16\%, and Audio-Text retrieval by 6\% and 23\%.

\end{abstract}
\vspace{-0.15in}

%% file: texfiles/introduction2.tex
\vspace{-0.05in}
\section{Introduction}\label{sec:intro}
\vspace{-0.05in}

\begin{table*}[]
\footnotesize
\center
\begin{tabular}{ccccccccccc} \hline
Dataset & Source & Language & Dur. (h) & \ audios & captions & \makecell{max \\ dur.(s)} & \makecell{avg \\ dur.(s)} & \makecell{max \\ words} & \makecell{avg \\ words} & Text Source \\ \hline
AudioCaps \cite{audiocaps} & YouTube & English & 135.01 & 50535 & 55512 & 10.08 & 9.84 & 52 & 8.80 & Human captions \\
Clotho \cite{clotho} & FreeSound & English  & 37.05 & 5929 & 29645 & 30.00 & 22.50 & 21 & 11.34 & Human captions \\
MACS \cite{macs} & TUT & English &  11.88 & 3930 & 17275 & 10.88 & 10.88 & 40 & 9.24 & Human annotated \\ 
SoundDescs \cite{sounddescs} & BBC & English  & 1060.40  & 32979 & 32979 & 4475.38 & 115.75 & 65 & 15.28 & Descriptions \\ \hline
WavText5K & \makecell{BigSoundBank \\ SoundBible}  & English & 25.48 & 4525 & 4348 & 2438.65 & 20.27 & 82 & 12.50 & Title, Descriptions \\ \hline
\end{tabular}
\caption{\label{table: dataset-description} The first three datasets have curated processes to annotate the captions, whereas SoundDescs and WavText5K use the free-form descriptions provided by the uploader of the audio recording.\vspace{-0.1in}}
\end{table*}


The audio-retrieval technology has numerous applications in search engines, anomaly detection, and audio and video editing. We will use audio-retrieval to refer to both text-audio and audio-text retrieval. The early works in audio-retrieval focused on using audio tags or events \cite{10.1145/1460096.1460115, Ikawa2018AcousticES, earlycrossmodal, elizalde2020never}. With availability of audio captioning datasets \cite{audiocaps,clotho,sounddescs}, the audio-retrieval task was expanded to include natural language descriptions as queries \cite{oncescu2021audio,xie2022dcase}. 
The difficulty of building a high recall audio-retrieval system is made evident from the mAP and recall metrics from the DCASE 2022 Task 6B baseline \cite{xie2022dcase}. In our first contribution, we focus on improving audio-retrieval systems training data. We introduce a new collection of web-crawled audio-text pairs. WavText5K consists of isolated audio events and their descriptions. For example, an isolated audio event description for a crow is: ``A single crow crying in the middle of the night". This description provides more information than prompts like ``this is a sound of crow" and yet at the same time is focused on one sound event of crow crying. We show how using WavText5K improves audio-retrieval performance on benchmark datasets.



The model architecture for audio retrieval \cite{sounddescs, mei2022metric, oncescu2021audio} consists of an audio encoder and a text encoder which learn a joint multimodal space. Current SOTA results \cite{mei2022metric} are obtained by training independent audio-retrieval models for each target dataset. However, audio-pretraining works \cite{elizalde2022clap, wav2clip, audioclip} show that large models trained on audio-text pairs can generalise to different tasks and domains. Particularly, a type of such model called CLAP \cite{elizalde2022clap} learns audio concepts from natural language and achieves SOTA in multiple domains. Therefore, we propose an architecture based on CLAP training that trains a single model and performs well on multiple audio-retrieval datasets. Also, the choice of the audio encoder has a significant impact on the audio-retrieval performance. Recent literature has shown audio transformers \cite{pann,gong21b_interspeech,gong2021ssast,htsat} work well on variety on downstream tasks. However, the transformer models can intake limited input patches and tokens. This is unfavourable for training audio-retrieval models which have to learn temporal dependencies over 20-30 seconds audio clips. So instead, our proposed architecture combines the audio transformers \cite{htsat, gong2021ssast} with CNN models \cite{pann} to understand variable-length audio scenes.  



In all, we make two contributions in this paper. First, we introduce a collection of audio-text pairs referred to as WavText5K. We show how WavText5K improves performance on audio-retrieval benchmarks.
Second, we propose an architecture for audio retrieval based on CLAP training and complementary audio encoders. The two contributions beat the audio retrieval SOTA for AudioCaps and Clotho on Text-Audio retrieval by a relative 2\% and 16\%, and Audio-Text retrieval by 6\% and 23\% respectively.

%% file: texfiles/newdataset.tex
\vspace{-0.05in}
\section{WavText5K}\label{sec:newdataset}
\vspace{-0.05in}
In this section, we introduce the WavText5K collection consisting of 4525 audios, 4348 descriptions, 4525 audio titles and 2058 tags. 
\vspace{-0.15in}
\subsection{Collecting audio-text pairs}\label{sec:datasetcollection}
\vspace{-0.05in}
The WavText5K\footnote{https://github.com/microsoft/WavText5K} data was sourced from two main website: BigSoundBank \footnote{https://bigsoundbank.com} and SoundBible \footnote{https://soundbible.com}. The website contains free to download, royalty-free library of sounds effects. The BigSoundBank website consists of sound effects in WAV, BFW, AIFF, MP3, OGG format with audio title and audio descriptions available. The SoundBible website consist of sound effects in WAV or MP3 with audio titles, descriptions, and tags available. BigSoundBank also has other metadata available like channels, conditions, sound type, bit depth etc, which we do not collect. The sampling rate of audio files vary and we resample all audio to 44.1 kHz. While collecting the audio, we encountered empty audio files, incorrect download links, and empty metadata. We removed those entries from the final collection. 
After removing, the dataset contains 4505 audios, 4348 descriptions, 4525 audio titles and 2058 tags.
\vspace{-0.15in}
\subsection{Data Analysis}\label{sec:dataanalysis}
\vspace{-0.05in}
WavText5K is compared with the existing audio-text datasets in Table \ref{table: dataset-description}. Though WavText5K focuses on isolated sound events and scenes, the average number of words in descriptions (12.5 words) is comparable to other datasets. Similarly, the average audio duration (20.27 seconds) is longer than datasets like AudioCaps and comparable to ClothoV2. WavText5K also has titles which can be used together with the descriptions to form captions (see Section \ref{subsec:ablation results}). 

\begin{table}[h]
\footnotesize
\center
\begin{tabular}{cl}
\hline
Dataset & Description \\ \hline
AudioCaps & \textit{A dog whimpers quietly.} \\ \hline
Clotho & \textit{\makecell[l]{A dog jumped into the water and swim and then came \\ out panting, it shook its body.}} \\\hline
MACS & \textit{There is traffic noise and a dog bark.} \\ \hline
SoundDescs & \textit{\makecell[l]{Mink Hunting - Huntsmen whistling and calling to dogs, \\ horns. Dogs in water and shaking off water.}} \\\hline
WavText5K & \textit{\makecell[l]{Behind a fence, two small dogs bark and grunt at 1 \\ meter from the microphone.}} \\ \hline
\end{tabular}
\caption{\label{table: Qualitative captions}
Randomly picked captions referring to the sound of a dog. \vspace{-0.1in}}
\end{table} 

\begin{figure}
    \centering
    \includegraphics[width=3.0in]{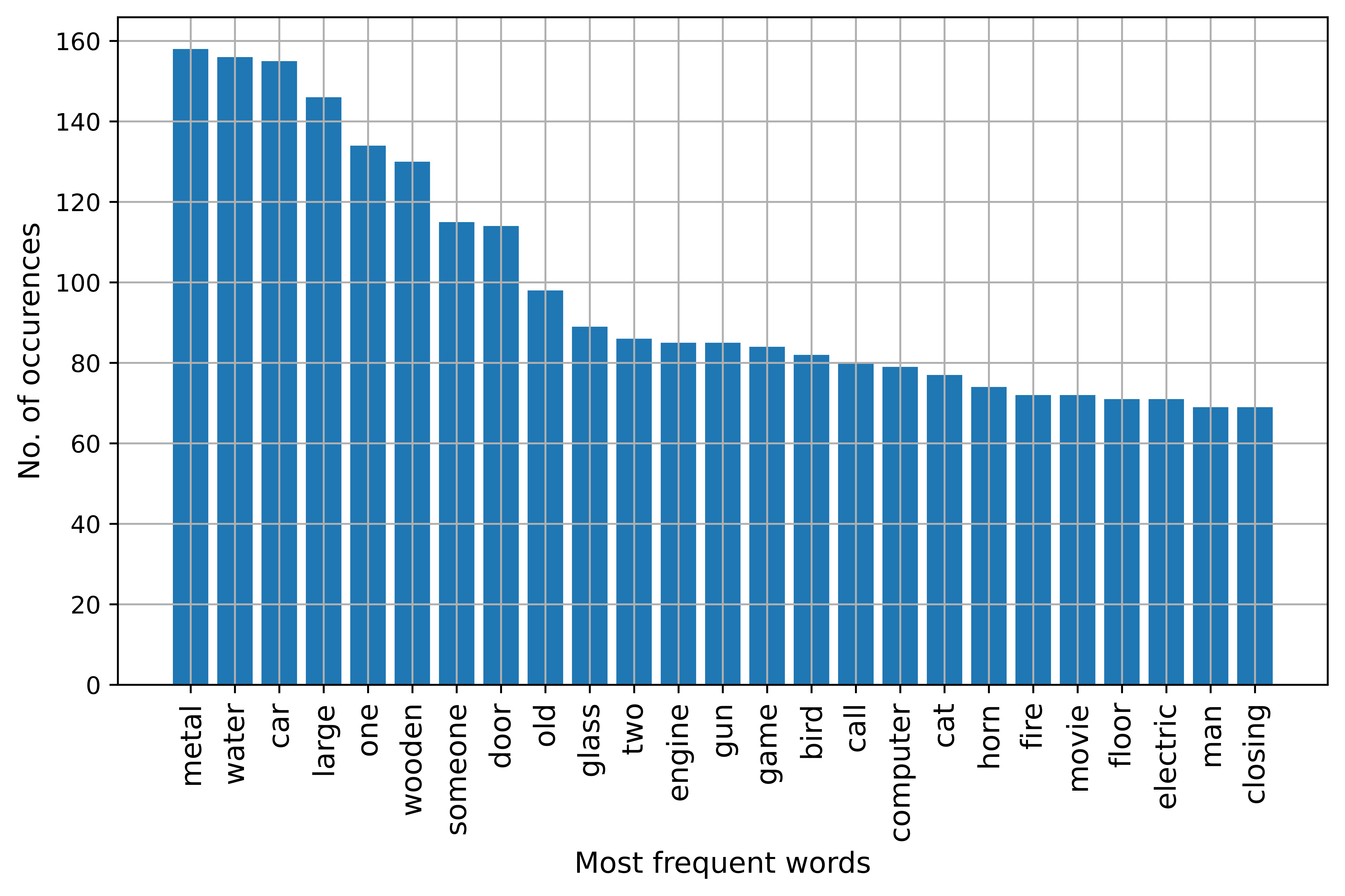}
    \caption{Most frequent words in WavText5K descriptions \vspace{-0.2in}}
    \label{fig:word frequency}
\end{figure}

We randomly sample caption from each dataset in Table \ref{table: dataset-description} and show them in Table \ref{table: Qualitative captions}. The Clotho captions contain multiple audio events like a dog jumping in the water (splash sound), swimming (paddling sound) and then panting and shaking its body. Comparatively, the WavText5K captions focus on the isolated event of dogs barking but provide much more context about the scene. To understand the audio events encountered in WavText5K, we plot the common words by frequency. The descriptions are filtered to remove punctuation, stop words, etc which have no semantic meaning. Then the filtered description is used to calculate the frequency of words and the top words are shown in Figure \ref{fig:word frequency}.



%% file: texfiles/architecture.tex
\vspace{-0.05in}
\section{Audio-Retrieval with contrastive learning}
\label{sec:architecture}
\vspace{-0.05in}

\begin{figure*}[ht]
   \centering
     \includegraphics[width=\textwidth,scale=0.9]{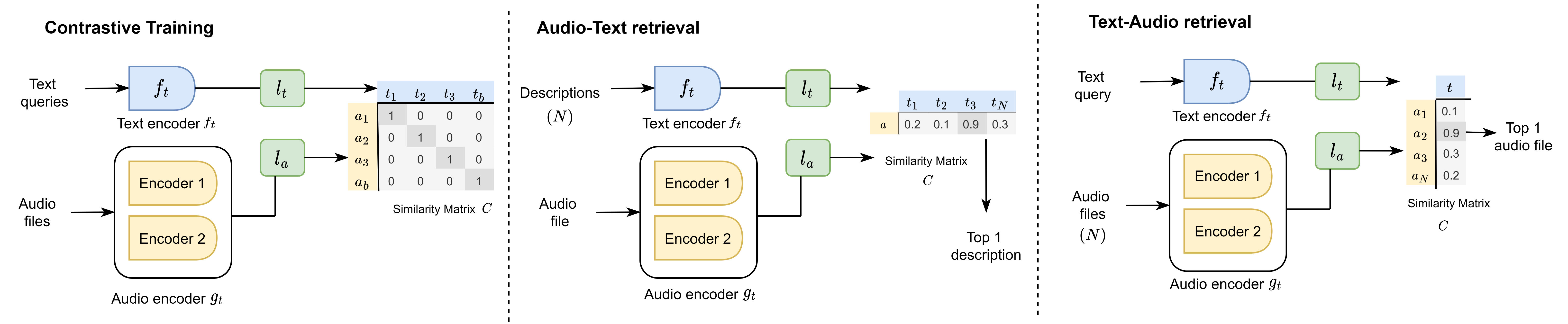}
     \caption{The model is trained on audio-text pairs using CLAP. At testing time, the trained encoders match the audio query to descriptions (audio-text retrieval) or text query to audio files (text-audio retrieval) in the database. \vspace{-0.08in}}
     \label{fig:architecture}
\end{figure*}

The CLAP model jointly trains audio and text encoder to learn a common multimodal space using contrastive learning \cite{elizalde2022clap}. The trained audio and text encoder are then later used to retrieve appropriate files for audio-text and text-audio retrieval. The proposed architecture is shown in Figure \ref{fig:architecture}.

\vspace{-0.15in}
\subsection{CLAP Training}\label{sec:clap}
\vspace{-0.05in}

Let the training data be $D = \{(a_i, t_i)\}_{i=1}^{i=N}$. Let $f(a)$ be the audio encoder and $g(t)$ be the text encoder which are learnable embedding functions. Here, the audio encoder $f(a)$ converts the raw audio into a log Mel spectrogram followed by a learnable embedding function. For a batch size of $b$:
\begin{equation}
    x_{a} = \{f(a_i)\}_{i=1}^{i=B} ; x_{t} = \{f(a_i)\}_{i=1}^{i=B}
\end{equation}
where $x_a\in \mathbb{R}^{b \times v}$ are the audio representations of dimension $v$, and $x_t \in \mathbb{R}^{b \times u}$ are the text representations of dimension $u$. The audio and text representation are brought into a common multi-modal space of dimension $d$ by independent linear projection layers $l_a(a)$, $l_t(t)$. This results in:
\begin{equation}
    \hat{x}_{a} = l_a(x_a); \hat{x}_{t} = l_t(x_t)
\end{equation}
where $\hat{x}_a\in \mathbb{R}^{b \times d}$ and $\hat{x}_t\in \mathbb{R}^{b \times d}$. Once both audio and text embeddings are in common embedding space, we can compare their similarity as:
\begin{equation}
    C = \tau*(\hat{x}_a \cdot \hat{x}_t^\top)
\end{equation}
where $\tau$ is a temperature parameter and the similarity matrix $C$ has $b$ correct pairs in the diagonal. To learn the embedding functions and projection layers we use symmetric cross-entropy loss ($\mathcal{L}$):
\begin{equation}
     \mathcal{L} = 0.5 * (\ell_{text}(C) + \ell_{audio}(C))
\end{equation}
where $\ell_{k} = \frac{1}{N}\sum_{i=0}^{N} \log diag (softmax(C))$ along text and audio axis respectively

\vspace{-0.15in}
\subsection{Audio Retrieval}
\label{sec:audio-retrieval}
\vspace{-0.05in}

After CLAP training, the model is used for audio-retrieval as shown in Figure \ref{fig:architecture}. The audio file(s) are embedded by audio encoder $f_t$ and the descriptions by text encoder $g_t$. This is followed by independently projecting ($l_t$, $l_a$) the embeddings into common multimodal space and computing the similarity matrix between the audio and text embeddings. This similarity matrix is represented as $C$ in Figure \ref{fig:architecture}. For audio-text retrieval, top-N descriptions are computed by picking the descriptions corresponding to the top N values in similarity matrix $C$. Similarly, for text-audio retrieval, top-N audios are computed by picking the audios corresponding to the top N values in similarity matrix $C$.

\vspace{-0.15in}
\subsection{Audio and Text encoder}\label{sec:encoders}
\vspace{-0.05in}

The CLAP model \cite{elizalde2022clap} used Cnn14 \cite{pann} as audio encoder and BERT as text encoder. There have been recent advances in audio transformer models \cite{gong21b_interspeech,gong2021ssast,htsat} which show comparable or better performance than CNN models. However, the transformer models can intake limited input patches and tokens. For example, the HTSAT is trained with 10 seconds audio clips. This is unfavourable for audio-text pair training where the audio concepts and complex descriptions have temporal dependencies which evolve over 20-30 seconds audio clips. So instead, we propose using a combination of CNN14 and HTSAT as the audio encoder:
\begin{equation}
    f(a) = \text{Concat}(CNN14(a), HTSAT(a))
\end{equation}
In section \ref{subsec:new dataset and architecture results}, we show that such complementary audio encoder improves audio-retrieval performance and beats SOTA. For text encoder, we use RoBERTa \cite{liu2019roberta} instead of BERT which is a more robust text encoder. We leave dynamic methods of combining audio embeddings like attention mechanisms \cite{deshmukh2021improving, deshmukh2020multi} for future work.

%% file: texfiles/experiments.tex
\vspace{-0.05in}
\section{Experiments}\label{sec:experiments}
\vspace{-0.15in}
\subsection{Datasets}\label{sec:training datasets}
\vspace{-0.05in}

We use Clotho \cite{clotho}, AudioCaps \cite{audiocaps}, and proposed WavText5K dataset for CLAP training. 
For Clotho train set, we use 4,884 audios and 24,420 captions. For AudioCaps train set, after downloading from YouTube, we end up with 40,582 audio files and captions. To have same number of audios and captions to form pairs, we repeat the audio files which have multiple captions. Therefore, the training data consists of 65,002 audio-text pairs. From WavText5K, we use 4348 audio-text pairs which have both description and title. For benchmarking, we use the test set of Clotho and test set of AudioCaps, where both of the sets are annotated by humans.

\begin{table*}[]
\footnotesize
\center
\begin{tabular}{lccccccc|ccccc}
\hline
 &  &  & \multicolumn{5}{c}{Text-Audio Retrieval $\uparrow$} & \multicolumn{5}{c}{Audio-Text Retrieval $\uparrow$} \\ \cline{4-13}
 \makecell{Exp - Audio \\ encoder} &\makecell{Training\\dataset}& \makecell{Retrieval\\dataset} & mAP@10  & R@1 & R@5 & R@10 & R@50 & mAP@10 & R@1 & R@5 & R@10 & R@50 \\ \hline 
Benchmark \cite{mei2022metric} &AudioCaps& AudioCaps & - & 33.9 & 69.7 & 82.6 & - & - & 39.4 & 72.0 & 83.9 & - \\ 
\textit{A} - CNN &AC, Cl & AudioCaps & 45.28 & 33.07 & 67.30 & 80.30 & 95.74 & 25.65 & 39.76 & 73.72 & 84.64 & 97.04 \\
\textit{B} - CNN &AC, Cl, WT5K& AudioCaps & 46.57 & 33.42 & 68.00 & 79.95 & 96.42 & 26.30 & 38.68 & 70.35 & 84.1 & 97.44 \\
\textit{C} - HTSAT &AC, Cl, WT5K& AudioCaps & 46.33 & 34.07 & 66.90 & 79.81 & 95.36 & 26.71 & 40.84 & 72.77 & 84.36 & 97.3 \\
\textit{D} - CNN-HTSAT & AC, Cl,WT5K& AudioCaps & \textbf{49.45} & \textbf{34.69} & \textbf{70.22} & \textbf{82.0} & \textbf{97.28} & \textbf{30.81} & \textbf{41.91} & \textbf{73.18} & \textbf{84.64} & \textbf{97.71} \\ \hline
Benchmark \cite{mei2022metric} &Clotho& Clotho & - & 14.4 & 36.6 & 49.9 & - & - & 16.2 & 37.5 & 50.2 & - \\ 
\textit{A} - CNN &AC, Cl& Clotho & 24.74 & 15.79 & 36.78 & 49.93 & 80.75 & 12.41 & 17.42 & 40.57 & 54.26 & 82.68 \\
\textit{B} - CNN &AC, Cl, WT5K& Clotho & 25.85 & 16.48 & 39.58 & 52.46 & 82.0 & 12.87 & 18.47 & 44.02 & 57.51 & 86.03 \\
\textit{C} - HTSAT &AC, Cl, WT5K& Clotho & 22.62 & 14.24 & 36.11 & 49.29 & 82.47 & 10.15 & 16.36 & 38.37 & 50.43 & 81.15 \\
\textit{D} - CNN+HTSAT &AC, Cl, WT5K& Clotho & \textbf{27.12} & \textbf{16.75} & \textbf{41.09} & \textbf{54.07} & \textbf{83.79} & \textbf{13.65} & \textbf{20.0} & \textbf{44.88} & \textbf{58.66} & \textbf{87.65} \\ \hline
\end{tabular}
\caption{\label{table: audio retrieval results}
\textit{Exp D} is our proposed model with CLAP pretraining on WavText5K (WT5K), AudioCaps (AC), and Clotho (CL), and achieves state of the art outperforming the literature~\cite{mei2022metric}.\vspace{-0.1in}}
\end{table*}

\vspace{-0.15in}
\subsection{Experimental setups}\label{sec:experimental setup}
\vspace{-0.05in}

The audio files are resampled to 44.1 kHz and represented by log Mel spectrogram. The log Mel spectrogram is constructed with a hop size of 320, a window size of 1024, and 64 Mel bins in the range of 50-8000 Hz. Each audio file is randomly truncated to 20 secs for Cnn14 and 10 secs for HTSAT. 
The audio-text pairs are randomly sampled to form batches during training. The projection dimension for CLAP is set to be 1024 and the temperate $\tau$ is initialised to 0.007. We use Adam Optimiser \cite{adam} with the learning rate of $10^{-4}$ and is reduced by a factor of 0.1 every 20 epochs for a total of 45 epochs. The model is trained on 8 GPUs with a batch size of 128.  

%% file: texfiles/results.tex
\vspace{-0.05in}
\section{Results and Discussion}\label{sec:results}
\vspace{-0.05in}

We compare our results against the best in the literature \cite{mei2022metric} for both, ClothoV2 and AudioCaps, in Table \ref{table: audio retrieval results}. First, we compare CLAP baseline against the literature benchmark in Section \ref{subsec:clap baseline}. Second, we show the effect of adding WavText5K and proposed architecture performance in Section \ref{subsec:new dataset and architecture results}. Lastly, Section \ref{subsec:ablation results} focuses on ablation study for constructing captions for WavText5K.


\vspace{-0.15in}
\subsection{CLAP baseline}\label{subsec:clap baseline}
\vspace{-0.05in}

\textit{Exp A} is our baseline with CLAP model using Cnn14 as the audio encoder and RoBERTa as the text encoder. For T-A retrieval, the baseline experiment outperforms the ClothoV2 benchmark by 9.6\% and underperforms the AudioCaps benchmark by 2.5\% on R@1. For A-T, the baseline outperforms on ClothoV2 and AudioCaps by 7.5\% and 0.9\% respectively. As noted in \cite{sounddescs}, the Clotho dataset is particularly more challenging than AudioCaps due to its varied audio content distributed in 10-30 seconds audio files. This is unlike AudioCaps which is limited in temporal dependencies between audio events to 10 seconds. 
Therefore, the 9.6\% and 7.5\% improvement in T-A and A-T over the Clotho benchmark indicates that the CLAP training is better at learning long-term temporal dependencies compared to the previous metric learning objectives \cite{mei2022metric}.

\vspace{-0.15in}
\subsection{WavText5K and proposed architecture}\label{subsec:new dataset and architecture results}
\vspace{-0.05in}

\textit{Exp B} includes WavText5K in the training datasets. For each audio-text pair, we construct the caption as  \textit{``\{title\}. \{description\}"}. By adding WavText5K, the T-A R@1 improves over \textit{Exp A} by 1\% on AudioCaps and 4.4\% on Clotho.
This might be explained because WavText5K tends to have more isolated audio events and less complex acoustic content. 
To study effect of combining multiple training datasets, we also added SoundDescs \cite{sounddescs} and MACS \cite{macs} from Table~\ref{table: dataset-description}  to CLAP training of \textit{Exp B}. However, the mAP and R@1 performance decreased. Further studies are needed to determine how to best integrate MACS and SoundDescs.  


\textit{Exp C} uses HTSAT audio encoder and RoBERTa text encoder. Compared to Exp B, the performance increases on AudioCaps and decreases on Clotho. The HTSAT model has a limit on input patches it can process, which translates to 10 seconds of 32 kHz audio. Compared to HTSAT, Cnn14 can process and operate on 20 seconds of audio or more. Hence, Cnn14 is better suited for compound queries and audio scenes evolving over a longer duration like in the Clotho dataset. However, HTSAT improves performance on AudioCaps which is restricted to 10 seconds clips. 

In \textit{Exp D}, the proposed architecture (described in Section \ref{sec:encoders}) combines Cnn14 and HTSAT encoder, and is jointly trained with text encoder RoBERTa. In all, our two contributions of WavText5K and proposed architecture improves T-A R@1 over literature benchmark by 2.3\% and 16.3\% on AudioCaps and Clotho dataset respectively. Similarly, for A-T R@1, we see an improvement of 6.4\% and 23.5\% on AudioCaps and Clotho respectively.

\vspace{-0.15in}
\subsection{Caption construction}\label{subsec:ablation results}
\vspace{-0.05in}

We performed ablation study to understand the effect of caption construction using \textit{``\{title\}"}  and \textit{``\{description\}"}. The results are in Table \ref{table: ablation results}. On Clotho, using caption as \textit{``\{title\}. \{description\}"} provides about 3.5\% and 4.5\% improvement on T-A and A-T R@1 metrics against using only \textit{``\{description\}"} as caption. On AudioCaps, we observed mixed results where performance decreases by 0.2\% on T-A retrieval and increases on A-T retrieval by 4.4\%. This result indicates that the titles in WavText5K do have some additional information over description. However, the improvement in audio-retrieval performance is not as strong.

\begin{table}[]
\footnotesize
\center
\begin{tabular}{cccc|cc}
\hline
& & \multicolumn{2}{c}{T-A Retrieval $\uparrow$} & \multicolumn{2}{c}{A-T Retrieval $\uparrow$} \\ \cline{3-6}
& \makecell{Retrieval\\dataset} & mAP@10 & R@1 & mAP@10 & R@1 \\ \hline
\textit{Desc.} & AudioCaps & \textbf{49.54} & \textbf{34.77} & 30.48 & 40.16 \\
\textit{Desc. + Title} & AudioCaps & 49.45 & 34.69 & \textbf{30.81} & \textbf{41.91} \\ \hline
\textit{Desc.} & Clotho & 26.15 & 16.19 & 12.77 & 19.14 \\
\textit{Desc. + Title} & Clotho & \textbf{27.12} & \textbf{16.75} & \textbf{13.65} & \textbf{20.0} \\ \hline
\end{tabular}
\caption{\label{table: ablation results}
Constructing the caption with the title and the description of WT5K results in better retrieval performance. \vspace{-0.1in}}
\end{table}

%% file: texfiles/conclusions.tex
\vspace{-0.05in}
\section{Conclusion}\label{sec:conclusions}
\vspace{-0.05in}
We introduced WavText5K consisting of audio-text pairs with isolated audio events and descriptions. We proposed an architecture for retrieval which used symmetric cross entropy as objective function and combines the latest advancements in audio transformers with CNN models. In all, by training the proposed architecture with WavText5K in training data, beats both the Text-Audio and Audio-Text benchmarks on AudioCaps and Clotho.


